# Impact of dislocation densities on the microscale strength of single-crystal strontium titanate

Jiawen Zhang[1], Xufei Fang[2*], Wenjun Lu[1*]

[1]Department of Mechanical and Energy Engineering, Southern University of Science and Technology, Shenzhen, 518055, China

[2]Institute for Applied Materials, Karlsruhe Institute of Technology, 76131 Karlsruhe, Germany

*Corresponding authors: xufei.fang@kit.edu (XF); luwj@sustech.edu.cn (WL)

**Abstract**

Dislocations in ceramics at room temperature are attracting increasing research interest. Dislocations may bring a new perspective for tuning physical and mechanical properties in advanced ceramics. Here, we investigate the dislocation density dependent micromechanical properties of single-crystal $SrTiO_3$ by tuning the dislocation densities (from ~$10^{10}$ $m^{-2}$ up to ~$10^{14}$ $m^{-2}$). Using micropillar compression tests, we find the samples exhibit a transition from brittle fracture (if no dislocation is present in the pillars) to plastic yield (with pre-engineered dislocations in the pillars). While within the regime of plastic deformation, the yield strength and plastic flow behavior exhibit a strong dependence on the dislocation density. The yield strength first decreases and then increases with the increase of dislocation densities. Detailed examination via post-mortem transmission electron microscopy reveals a complex evolution of dislocation structure, highlighting the critical role played by dislocations in regulating the brittle/ductile behavior in $SrTiO_3$ at room temperature. Our findings shed new light on dislocation-mediated mechanical properties in ceramics and may provide designing guidelines for the prospective dislocation-based devices.

# 1. Introduction

Functional ceramics are widely used for energy storage, electronic devices, and catalysis [1–3]. However, their applications are often limited by the intrinsic brittleness of ceramics due to the strong ionic and/or covalent bonding. Continuous efforts have been made over decades to achieve plastically deformable ceramics but yielded little success except at small scales. Recent studies demonstrated that bond switching at coherent interfaces facilitates stress-induced phase transformation in $Si_3N_4$ ceramic and enables 20% plastic deformation at room temperature but is limited to the nano-/microscale [4]. Additionally, by constructing a metal-ceramic interface structure between Mo and $La_2O_3$, the dislocation transfer (borrowed-dislocations) is realized in the $La_2O_3$, achieving ~40% tensile plastic deformation during *in situ* small scale testing [5]. Another material system that exhibits room-temperature plasticity is $ZrO_2$-based ceramics, owing to the phase transformation upon stress activation [6]. Nevertheless, dislocation-mediated plastically deformable ceramics, particularly at room temperature and at macroscale, remain largely unattainable except for alkali halides e.g. LiF [7], MgO [8] with rock-salt structure, and perovskite oxides $SrTiO_3$ [9], $KNbO_3$ [10], and $KTaO_3$ [11], all experimentally verified by bulk compression.

To achieve dislocation plasticity in ceramics at room temperature, crack formation must be avoided by circumventing dislocation nucleation, which requires much higher shear stress. Mechanical deformation is one of the most common methods for engineering dislocations [12]. This is often not achieved, particularly for bulk ceramic samples, before the load exceeds the fracture strength of the samples [12]. To avoid cracking while promoting dislocation plasticity, it is common to adopt high-temperature deformation, where thermal activation facilitates dislocation generation and motion. For instance, single-crystal $TiO_2$ and $Al_2O_3$ exhibit a room-temperature compressive strain of approximately 10% and 7.5%, respectively, after preloading and deformation at elevated temperatures, correlated with the enriched dislocation density induced during pre-deformation at high temperature [13].

Alternative to high-temperature deformation and conventional bulk deformation for dislocation imprinting [14, 15], room-temperature near-surface processing techniques such as surface grinding [16], nanoindentation [17], and cyclic Brinell indentation [18] have proven feasible for introducing dislocations. First of all, surface grinding creates dislocations with a gradient in the near-surface region of 1~5 μm in depth, producing a dislocation density ranging from ~$10^{13}$ m$^{-2}$ and higher as in the case of $SrTiO_3$ [16, 19]. However, precise control of dislocations using this method remains unfeasible for surface grinding, and the depth of dislocation-rich regions is limited to the skin area of the sample. Second, nanoindentation offers a solution for applying localized high shear stress to nucleate dislocations within a small volume. Many studies have explored nanoindentation-induced dislocations in various ceramics at room temperature, such as $BaTiO_3$ [20], ZnO [21], MgO [22], and $SrTiO_3$ [23, 24]. Often, spherical indenters are preferred to minimize stress concentration at the sharp edges as in Berkovich indenter. In this regard, understanding the competition between dislocation activation and crack formation has been a key focus, particularly concerning the critical indenter tip radius for various oxides [17]. However, the drawback of nanoindentation tests lies in their very localized plastic zone and dislocation distribution, typically spanning only several microns in depth and width. To overcome these limitations to further increase the plastic zone size as well as the dislocation density, Okafor et al. demonstrated a cyclic indentation method, using a millimeter sized Brinell indenter, to generate a large and crack-free plastic zone in $SrTiO_3$ [18]. By adjusting the cyclic loading number, tunable dislocation densities have been achieved ranging from $\sim 10^{12}$ m$^{-2}$ to $\sim 10^{14}$ m$^{-2}$ in $SrTiO_3$ and $KNbO_3$, with plastic zone size of ~150 μm in depth and width [18, 25].

The successful engineering of dislocations into ceramics has led to studies of the dislocation-tuned mechanical properties. For instance, Okafor et al. and Preuß et al. evaluated the Vickers hardness as a function of the dislocation density, and observed an increase in the micro-hardness as the dislocation density increases [18, 26]. As indentation tests are accompanied with a complicated stressed field, to avoid this complexity, Fang et al. [19] used the surface grinding technique described above and tested two extreme cases: pillars with no dislocations and with a mechanically seeded dislocation density of $\sim 10^{14}$ m$^{-2}$ in the skin area of single-crystal $SrTiO_3$. It was demonstrated that micropillars (with a diameter of ~4 μm) without dislocations fractured in a brittle manner, while the pillars with mechanically seeded dislocations deformed beyond 20% plastic strain.

However, the two extreme conditions probed by Fang et al. [19] leave an open, pertinent question: how would the dislocations with various densities affect the strength of ceramics, provided that the dislocation densities can be tuned? Considering the hardness and yield strength relation in metals [27], one might expect the yield strength to increase with increasing dislocation density in ceramics. To address this question, we use single-crystal $SrTiO_3$ as a model material to evaluate the dislocation density dependent yield strength. $SrTiO_3$ is the first perovskite oxide reported with room-temperature bulk plasticity [9], and tunable dislocation densities [18]. The choice of single crystals allows us to examine the impact of dislocations on the mechanical properties without the complexity caused by grain boundaries.

Here, we adopt the Brinell cyclic scratching method [28] to generate large, crack-free plastic zones with tunable dislocation densities (from $\sim 10^{10}$ m$^{-2}$ up to $\sim 10^{14}$ m$^{-2}$) in the single crystal $SrTiO_3$. The length of the plastic zone (scratch track) extends to several mm depending on the sample size to allow for sufficient space for testing at a later stage. Micropillars were then prepared using focused ion beam (FIB) within the scratch tracks and tested during compression to directly correlate dislocation density with the micromechanical response. Post-mortem TEM (transmission electron microscopy) was used to reveal the dislocation structure inside the micropillars to shed light on the deformation behavior.

## 2. Materials and Methods

### 2.1. Materials preparation

Undoped, single-side polished (001) SrTiO$_3$ single crystals (Hefei Ruijing Optoelectronics Technology Co., Ltd., Anhui, China) with dimensions of 5×5×1 mm$^3$ were used. The reference crystals have a low initial dislocation density (~10$^{10}$ m$^{-2}$, with an average dislocation spacing of ~10 μm). Dislocations were introduced analogously to a previously established procedure [28], using a wear testing machine (Rtec, MFT2000, UK) equipped with a spherical indenter (Al$_2$O$_3$ ruby sphere, 3 mm in diameter) for surface scratching. The SrTiO$_3$ sample was affixed to a stainless-steel plate using crystal bond and positioned on a mobile stage capable of precise movements along the *x* and *y* axes (accuracy of 0.01 mm). Lubricant (polydimethylsiloxane methyl silicone oil) was used to minimize wear-induced crack initiation. The scratching was performed under a normal load of 9.5 N, with a lateral scratching speed of 0.2 mm/s and a wear track length between 1 and 2 mm, along the <100> direction. Scratch cycles were conducted at intervals of 1×, 5×, 20×, and 50× (with × representing the number of passes). Before the scratching started, the spherical indenter was gently brought into contact with the sample surface, with the load gradually increasing to 9.5 N over 10 seconds. During the scratching process, the load was maintained at a constant 9.5 N.

### 2.2. Microstructure characterization

Before scratching, a scanning electron microscopy (SEM, Apreo2 S Lovac, USA) equipped with an electron backscatter diffraction (EBSD) detector was used to confirm the crystal surface orientation. The operating voltage, current, and step size were 20 keV, 3.2 nA, and 100 nm, respectively. The surface morphology of the plastic zone created by cyclic scratching was characterized using an optical microscope (HVS-1000TM/LCD, Shanghai Optical Instrument Factory, China). Additionally, a three-dimensional optical profilometer (Countor GT K, Bruker, USA) was used to examine the changes in depth and width caused by different scratching passes.

After scratching, TEM specimens were prepared using a dual-beam focused ion beam (FIB) in an SEM (Helios Nanolab 600i, FEI, Hillsboro, USA). The TEM samples were lifted out both along and perpendicular to the scratching direction from the center of the wear track, allowing for better visualization of the dislocation structures from different observation perspectives. Scanning TEM (STEM) images were captured using a TEM instrument (FEI Talos F200X G2, Thermo Fisher Scientific, USA) operating at 200 kV. In the annual dark field (ADF)-STEM images, a probe semi-convergence angle of 10.5 mrad and inner and outer semi-collection angles of 23-55 mrad were used. For the annual bright field (ABF)-STEM images, inner and outer semi-collection angles of 12-20 mrad were used. 3D reconstruction of the spatial distribution of the dislocations was conducted using the region after 20× scratching, the lamella was lifted along the scratching direction using FIB and milling to ~200 nm in thickness. A tilt series of the ABF-STEM images of the dislocation structure were collected with a high-angle tomograph holder (Model 2020, Fischione, USA) in the TEM with tilting angles ranging from -60° to +60° and step of 2° along the [001] direction. The tilting series images were then aligned using Inspect3D™ software (Thermo Fisher Scientific Inc.) and reconstructed by the simultaneous iterative reconstruction technique (SIRT). Avizo 3D™ software was then applied to display the 3D dislocation structures.

For dislocation core characterization, TEM lamella (~50 nm in thickness) was lifted out from the 20× scratching track along the scratching direction by FIB. High-angle annual dark field scanning TEM (HAADF-STEM) imaging was then performed in a double aberration-corrected transmission electron microscope (Titan Themis G2, FEI, Netherlands) operated at 300 kV. A probe semi-convergence angle of 17 mrad and inner and outer semi-collection angles ranging from 38 to 200 mrad were employed.

**2.3. Mechanical properties measurement**

By testing micropillars with different pre-engineered dislocation densities, we directly quantify how dislocations govern strength and plasticity at microscale. Micropillars, approximately 1 μm in diameter and 3 μm in height, were prepared by FIB with an ion beam voltage of 30 keV and milled sequentially with currents of 2.5 nA, 0.79 nA, 0.23 nA, and 80 pA. The final fine milling was adopted to minimize the possible damage to the micropillar surface. Compression tests were performed on micropillars from regions with scratching pass numbers of 0× (reference), 1×, 5×, 20×, and 50×, with each condition repeated 5 times for reproducibility. The pillar compression tests were performed on a nanoindentation instrument (Hysitron TI950, Bruker, USA) equipped with a diamond flat punch indenter (5 μm in diameter). The equipment operates with a displacement-controlled mode with a strain rate of ~$10^{-3}$ $s^{-1}$. Note that these pillar compression tests performed outside the SEM are useful to rule out any electron beam effect on the deformation behavior. After deformation, TEM samples were prepared from selected deformed micropillars to investigate the dislocation structure inside the pillars.

## 3. Results & Analyses

### 3.1. Dislocation generation and density tuning

Guided by the hypothesis that dislocation density dictates the mechanical behavior of single-crystal SrTiO$_3$, we first employ Brinell ball cyclic scratching to create large plastic zones with tunable dislocation densities (**Fig. 1**). Contrasting the smooth surface on the reference sample with (001) surface (0×, **Fig. 1a-b**), the surface slip traces (**Figs. 1c-f**) become denser and finer as the scratching number increases from 1×, 5×, 20×, up to 50× (note 1× means 1 pass). After 1× scratching, the horizontal and vertical slip traces (**Fig. 1c**) correspond to dislocations generated on the {110} planes, inclined 45° to the (001) surface being scratched. This observation is consistent with the previous report on room-temperature indentation in SrTiO$_3$ [18]. The cross-sectional profiles of the scratch tracks (**Fig. 1g**) show a maximum depth of ~400 nm (after 50× cycles) over a scratch track width of ~120 μm, suggesting a normally flat surface. Note that the dislocation penetration depth is larger than ~100 μm thanks to the large Brinell indenter, as revealed by the cross-section chemical etching (see Supplementary Material **Fig. S1**).

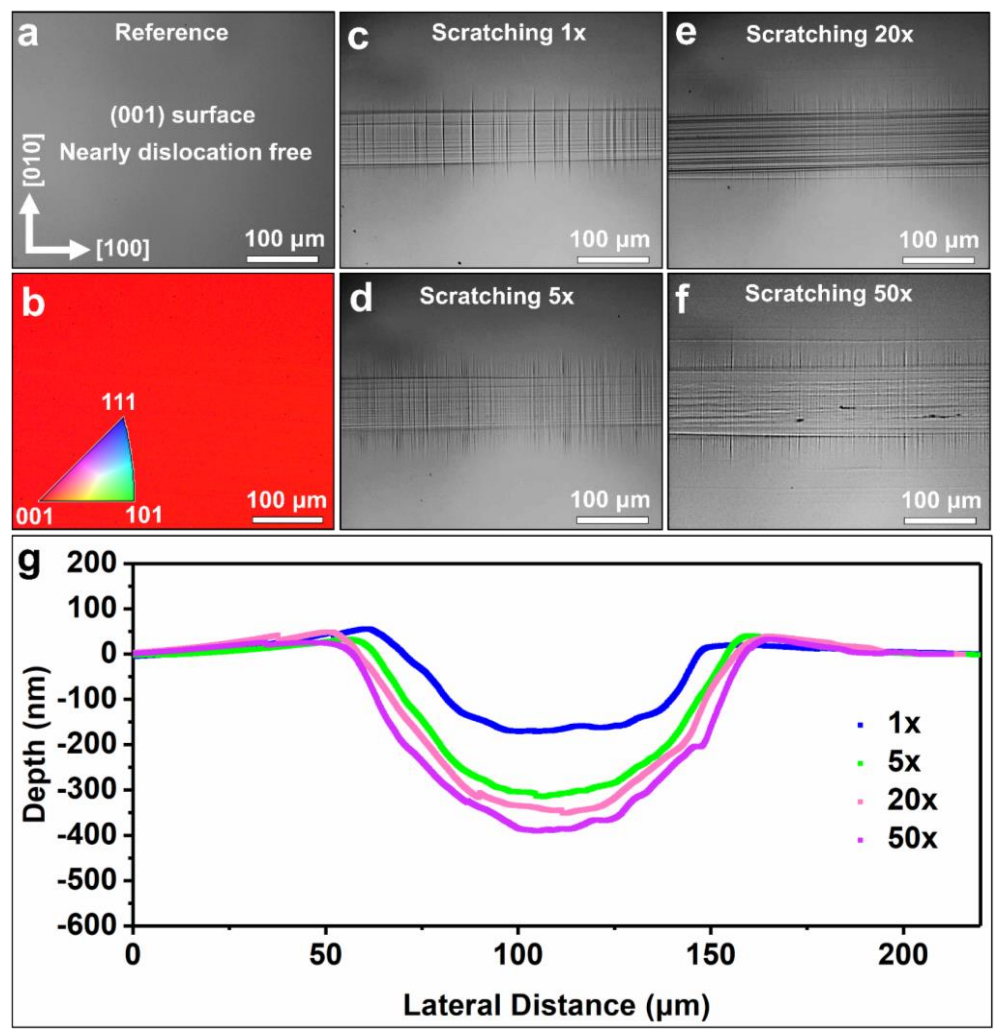

**Fig. 1.** Room-temperature cyclic scratching for mechanically imprinting dislocations into SrTiO$_3$. (a) Reference sample surface. (b) EBSD map confirming the (001) surface orientation. (c-f) Optical images of the sample surface after different scratching passes (1×, 5×, 20×, and 50×), revealing the surface slip traces. (g) Cross-section surface profile of the scratch tracks.

The slip trace density increase suggests a dislocation density increase, as visualized by TEM in **Figs. 2c-j**. To highlight the dislocation line contrast, all ADF-STEM images were collected along the [001] zone axis. TEM characterization along (**Figs. 2c-f**) and perpendicular (**Figs. 2g-j**) to the scratch direction demonstrates different spatial distribution patterns of the dislocations beneath the scratch tracks. The dislocation density was estimated using the line intercept method, which involves calculating the intersections in the TEM images [29]. After 1×, 5×, 20×, and 50× scratching, the dislocation densities are ~1.2×10$^{13}$ m$^{-2}$, 7.5×10$^{13}$ m$^{-2}$, ~1.6×10$^{14}$ m$^{-2}$, and ~3.3×10$^{14}$ m$^{-2}$, respectively. For comparison, the reference (0×) sample (**Fig. 2a**) is almost dislocation-free within the TEM sample. This is consistent with previous measurements made by the chemical etching method that revealed a dislocation density of about ~10$^{10}$ m$^{-2}$ in reference samples [23]. The dislocation density as a function of the scratching pass number is summarized in **Fig. 2b**. This cyclic scratching method proves to be an effective approach to significantly enhance dislocation multiplication, resulting in up to four orders of magnitude increase in the dislocation density.

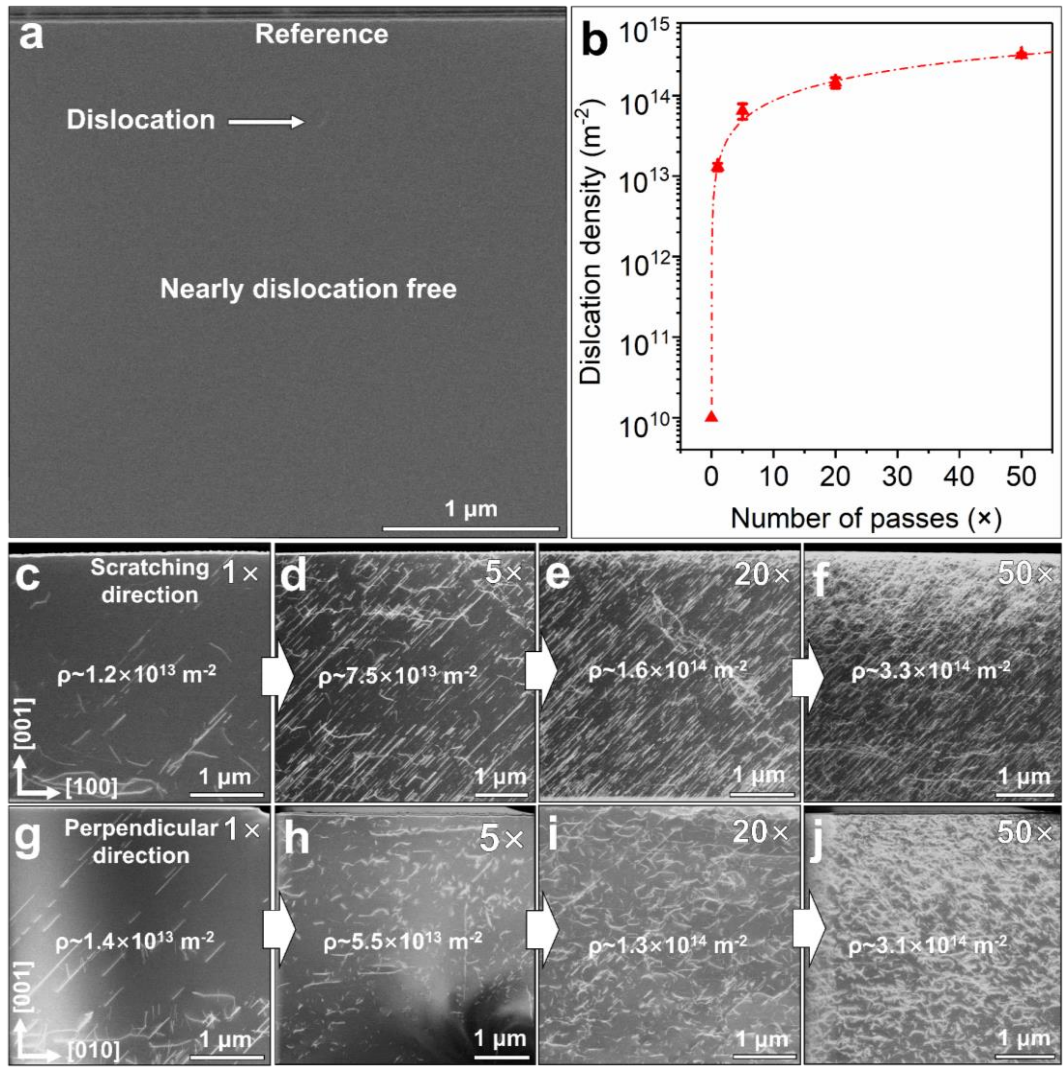

**Fig. 2.** (a) ADF-STEM image of a nearly dislocation-free region in the reference sample. One short dislocation line was indicated by the white arrow. (b) Overview of the dislocation densities varying as a function of the number of scratching passes. (c-f) A series of ADF-STEM images showing the distribution of dislocations within the samples with 1×, 5×, 20×, and 50× scratching pass numbers along the scratching direction (i.e., [010]). (g-j) Dislocation distribution inside the samples various pass numbers perpendicular to the scratching direction (i.e., [100]).

Besides the density change, the dislocation structures also vary with scratching passes. After 1× scratching, dislocations are generated on {110} planes, inclined 45° to the scratched (001) surface. After 5× cycles, interactions between the dislocations on (101) plane and those on (011) plane lead to the formation of 45° cross-hatched structures (**Fig. 2d**). With 20× passes, further interactions between dislocations from the (0-11) plane result in 90° cross-hatched structures and tangled-up dislocations (**Fig. 2e**). Finally, after 50× passes, the activation of multiple slip planes through dislocation multiplication significantly decreases the distance among dislocations, leading to more severe interactions and the formation of cross-hatched and cell-like structures (**Fig. 2f**), indicating dislocation cross-slip may have taken place [30]. Since screw-type dislocations are important for cross-slip to operate, detailed analyses were further performed to reveal the dislocation types (screw, edge, or mixed), as demonstrated in **Fig. 3** and Supplementary Materials **Table S1**. In **Fig. 3a**, after 1× scratching, dislocations 1-4 (**Fig. 3a**) are identified as screw type. After 5× scratching, dislocations 1-2 (**Fig. 3g**) parallel to the sample surface are identified as 35.26° mixed type, while the 45°-inclined dislocation 3 (**Fig. 3g**) is a screw type.

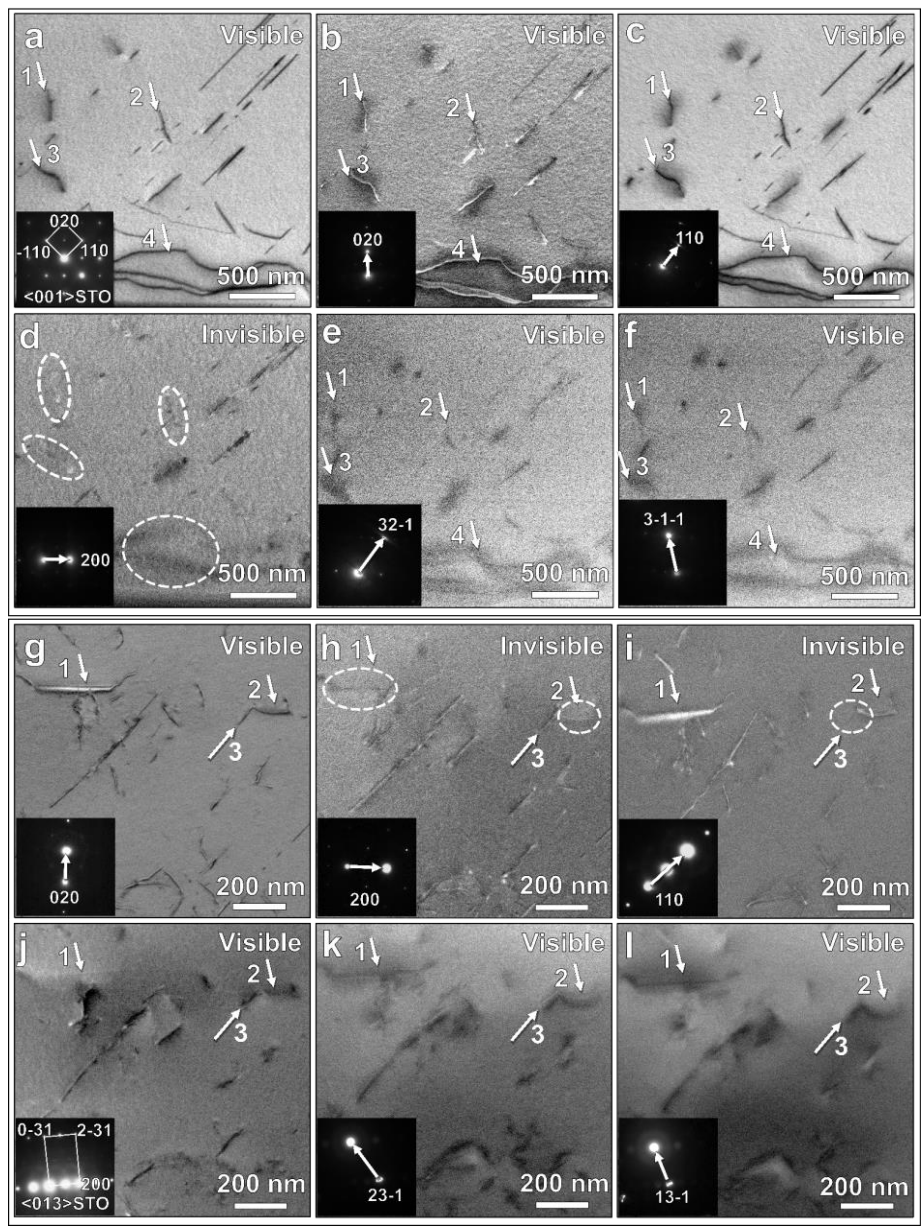

**Fig. 3.** Dislocation analysis after 1× and 5× scratching (with relatively lower density for easier analysis). Tilting series of the 1× scratching induced dislocations: (a) along <001> zone axis, (b) with **g**=(020), (c) with **g**=(110), (d) with **g**=(200), (e) with **g**=(32-1), and (f) with **g**=(3-1-1). Tilting series of the 5× scratching induced

dislocations: (g) with **g**=(020), (h) with **g**=(200), (i) with **g** =(110), (j) along <013> zone axis, (k) with **g** =(23-1), and (l) with **g**=(13-1).

The previous analyses are all in 2D. For a better understanding of the spatial distribution of dislocations, **Figure. 4** presents a 3D reconstruction of the dislocation networks after 20× scratching using ABF-STEM tomography. **Figs. 4a-d** and **4f** show a part of the tilt ABF-STEM images collected at 15°, 0°, -15°, 30° and -30°. The 3D reconstruction in **Fig. 4e** and tilt series of ±15° (**Figs. 4g** and **4i**) reveal a majority of forest-like dislocations lying on the (011) planes and (0-11) planes.

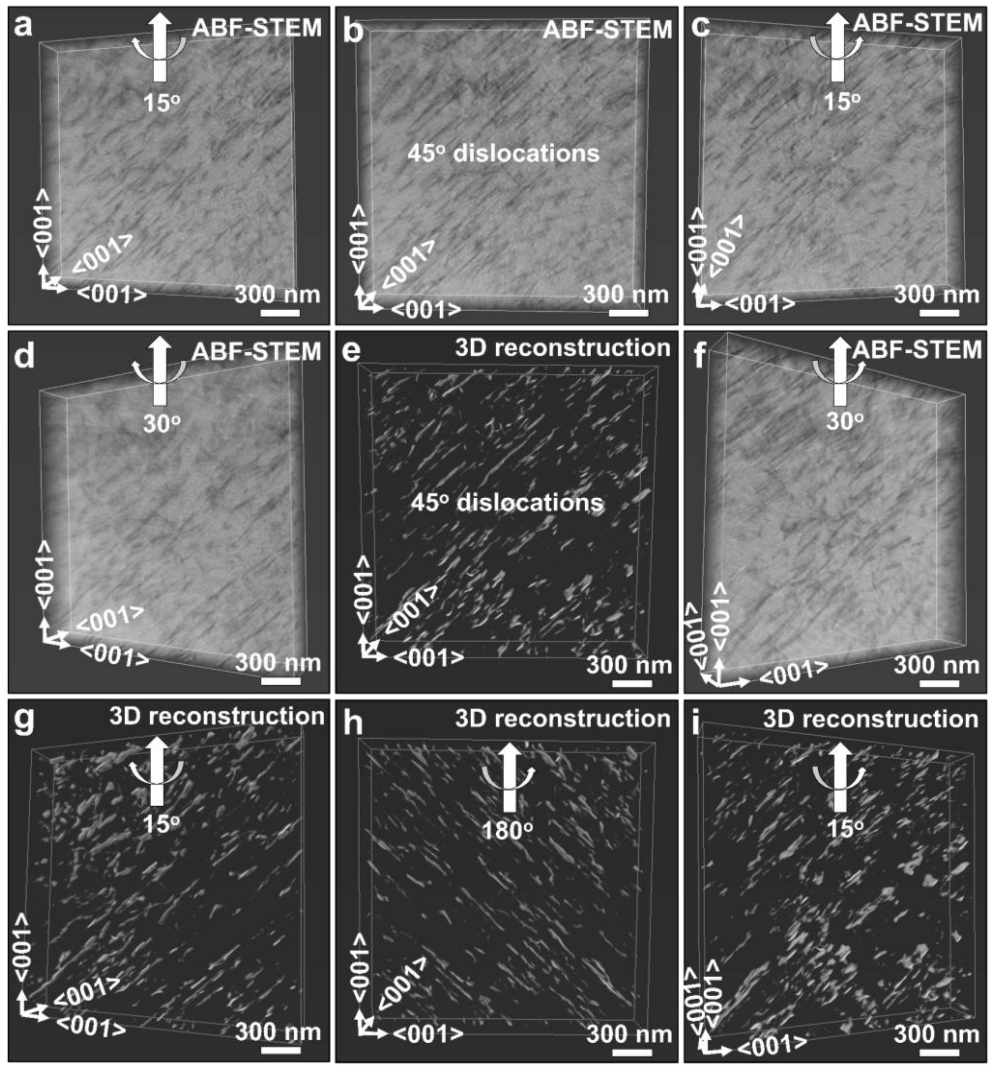

**Fig. 4.** ABF-STEM tomography of the dislocations (20× scratching) viewed at different tilting angles (indicated in each image). 3D reconstructed dislocations with the tilting angles: (e) 0°, (g) 15°, (h) 180°, and (i) -15°.

As dislocation core structures are believed to play a critical role in dislocation mechanics [31], we used HAADF-STEM imaging, coupled with fast Fourier transform (FFT) and inverse FFT (IFFT), to investigate the core structures of the dislocations (**Figs. 5a-b, d-e**, after 20× scratching). Geometric phase analysis (GPA) maps (**Figs. 5c, f**), generated using the FFT pattern, allow for quantifying the strain field around edge dislocations. Screw dislocations were not detected here due to the common challenge, as end-on imaging exhibits limited sensitivity to displacements along atomic columns [32]. Two selected *g* vectors, *g*= -1-10 and *g*=-200, marked with green circles in the FFT pattern, were used to compute the strain field in the *x* direction ($\varepsilon_{xx}$). The resulting strain maps in **Figs. 5c, f** revealed areas of tensile strain (red) and compressive strain (blue) around the edge dislocations. We identified a dislocation dipole consisting of two partial dislocations with opposite Burgers vectors (**Figs. 5a-b**). Such dipole had a separation of approximately 1.6 nm, spanning about three-unit cells. This dipole configuration has not been observed before in SrTiO$_3$, and was likely formed during the significant dislocation multiplication after 20× scratching, consistent with the role of dislocation dipoles as sources of dislocation multiplication [33]. Additionally, climb-dissociated dislocations with identical Burgers vectors were observed at the tail of the 45°-inclined dislocation line (**Fig. 5d**). The climb distance was about 2 nm, consistent with previous observation [31]. Both the dislocation dipoles and the climb-dissociated dislocations induced compressive and tensile stresses around the dislocation cores, as illustrated by the strain maps (**Figs. 5c, f**).

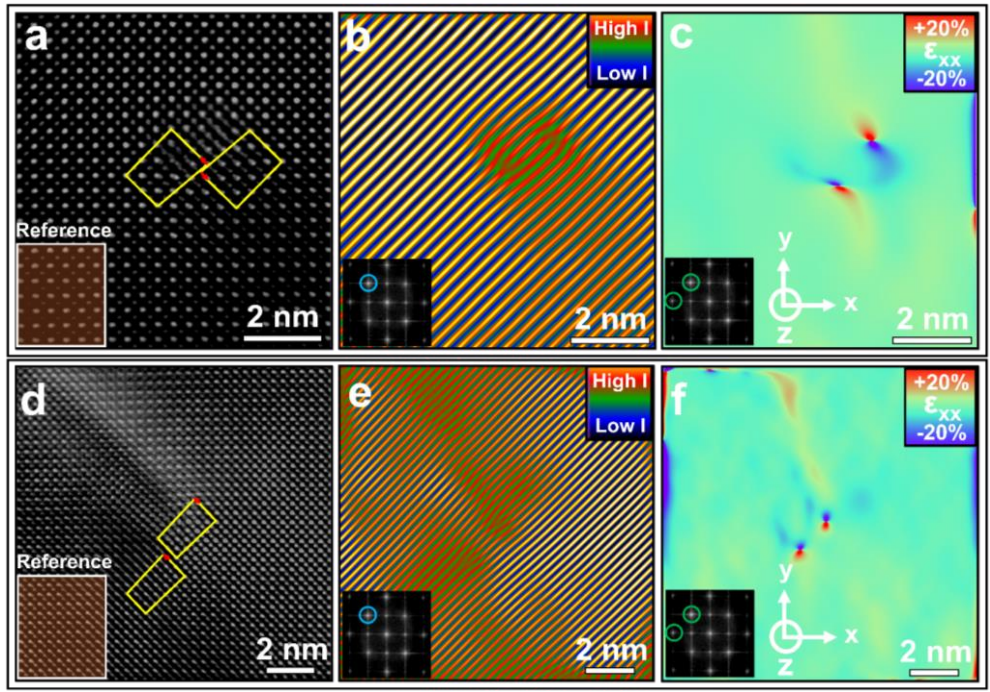

**Fig. 5.** Dislocation core structure analysis. (a) HAADF-STEM image shows a dislocation dipole with opposite Burgers vectors of 1/2<110> and -1/2<110> (indicated by red arrows) on the {110} plane. (b) Corresponding IFFT image for the atomic planes. The inset shows the FFT pattern. (c) Strain field around the dislocation dipoles in (a). The inset shows the FFT pattern, with two selected reflections (green circle) adopted for geometric phase

analysis (GPA). (d) HAADF-STEM image depicts two edge-type dislocations with Burgers vectors of 1/2<110>{110}. (e) The corresponding IFFT image reveals the atomic planes, demonstrating climb-dissociated partial dislocations. (f) Corresponding strain map for (d) around the dislocations. The region of observation is after 20× scratching.

It is worth noting that, contrasting the grain refinement observed in metallic materials due to scratching [34], even the severely deformed regions here in single-crystal $SrTiO_3$ remain in single-crystal form with such a high dislocation density after 50× scratching (**Fig. 2f, j**). This is confirmed by a SAED pattern (see Supplementary Materials **Fig. S2**) in TEM analysis. The tunable dislocation densities without the complexity of (sub)grain boundaries and the sufficiently large plastic zone size allow us to investigate the dislocation density-dependent strength.

### 3.2. Dislocation density dependent strength in micropillar compression

Subsequent micropillar compression within these regions reveals two key regimes: (i) dislocation-starved plasticity at low densities and (ii) dislocation entanglement induced hardening at high densities. This duality underscores the necessity of microscale testing to decouple dislocation effects from bulk-scale complexities. The compression tests on micropillars with different dislocation densities are presented in **Fig. 6**. In the reference sample (**Figs. 6a-d**), the micropillar undergoes pure elastic deformation up to ~5.3 GPa before a distinct displacement jump occurs, caused by brittle fracture. Although shear deformation occurs along the {110} plane under the fractured region, no noticeable slip bands were formed (**Fig. 6d**). With a dislocation density ~$7.5 \times 10^{13}$ $m^{-2}$ (5× scratching, **Figs. 6e-h**), the micropillar deforms by ~10%. The stress-strain curve (**Fig. 6e**) shows a flow stress of about 0.9 GPa ($\sigma_{0.2}$), with significant stress drops observed. The compressed pillar exhibits multiple coarse slip bands along the {110} planes (green arrows in **Figs. 6h**). These slip bands correspond to the large stress drops in the stress-strain curve (**Fig. 6e**), indicating dislocation avalanches during deformation. As the dislocation density continues to increase, representative stress-strain curves (**Fig. 6i** for 20× and **Fig. 6m** for 50×) and post-mortem SEM images demonstrate ~20% total strain without visible cracks. Interestingly, compared to the 5× pillar with a yield strength of ~0.9 GPa, the 20× and 50× pillars have higher yield strength of ~1 GPa and ~1.5 GPa, respectively.

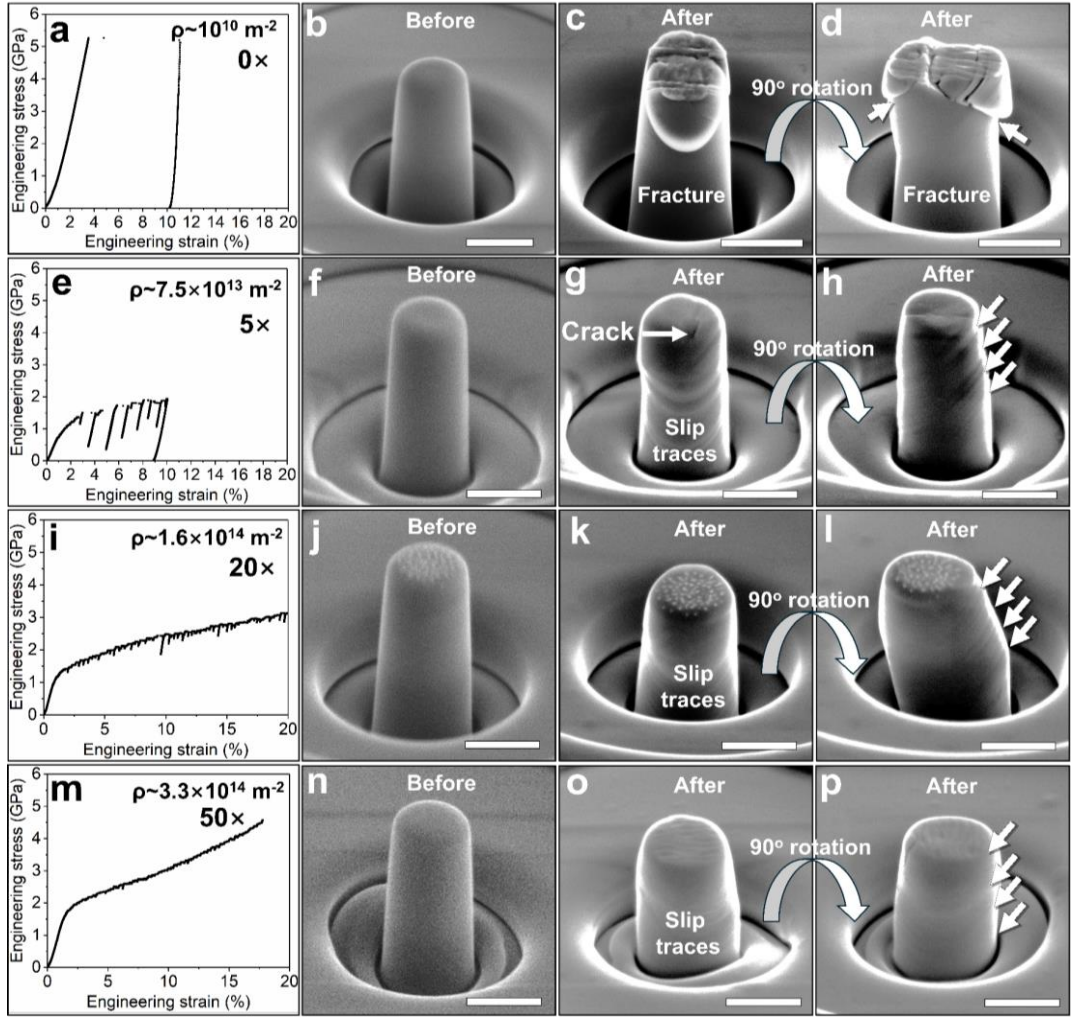

**Fig. 6.** Representative micropillar compression tests with varying dislocation densities: (a) Engineering stress-strain curve of a reference micropillar. (b) SEM image of the micropillar prior to compression. (c) and (d) SEM images showing brittle fracture at the top of the reference micropillar. (e, i, m) Engineering stress-strain curves for pillars after 5×, 20×, and 50× scratching. (f, j, n) SEM images for pillars after 5×, 20×, and 50× scratching, before compression. (g, h) SEM images revealing a crack formed at the pillar top and coarse slip traces for the 5× pillar. (k, l) SEM images showing much finer and denser slip traces in the deformed 20× pillar. (o, p) SEM images depicting smooth slip traces in the 50× scratching micropillar. The scale bar is 1 μm for all SEM images.

A noticeable feature in the stress-strain curves is the change in the magnitude of the stress drops with increasing scratching passes (hence increased dislocation densities, **Fig. 2b**). For the 5× pillar, the serrated stress-strain curve suggests pronounced dislocation avalanches during compression. Similar phenomena for large stress drops have been observed in pure Cu and Cu-Al single crystals in micropillar compression, with diameters ranging from 150 to 750 nm [35]. These abrupt stress drops are likely caused by the small pillar size and the low dislocation density, which facilitate the easier escape of the dislocations to the free surface of the pillars. In FCC metals, flow stress is influenced by pillar size, dislocation density, and dislocation distribution [36]. Generally, pillars with smaller diameters exhibit more pronounced stress drops (see further discussion and validation in **Sec. 4**), while higher dislocation densities result in denser and smaller stress drops [36], indicating weaker dislocation avalanche activities due to reduced distance and stronger hindering interactions among adjacent dislocations [37]. This is the case for the 20× and 50× SrTiO$_3$ pillars (**Fig. 6i, m**), where the serrations become weaker for the 20× pillar, and almost disappeared for the 50× pillar. This progression suggests a weakening of the dislocation avalanches in the micropillars with higher dislocation densities. For the 50× pillar, the homogeneously distributed dislocations (see **Figs. 2f, j**) lead to a continuous and smooth deformation. Overall, compared to the reference sample that exhibits fracture-dominated behavior, the improved plasticity in the 5×, 20×, and 50× scratched pillars can be attributed to easier dislocation motion and multiplication compared to dislocation nucleation which requires much higher shear stress [19].

Now we attend to the pillars in the 1× scratch track. These pillars (**Fig. 7**) exhibited a stochastic deformation behavior, namely, some of them fractured while some yielded. By checking the dislocation density after 1× scratching, this stochastic behavior can be understood: because the average dislocation spacing (~1 μm, see **Figs. 2c, g**) matches the diameter of the micropillars, those do not have pre-existing dislocations tend to fracture (as the reference sample **Figs. 6a-d**), whereas the ones with scratch-induced dislocations can deform plastically. These results emphasize the critical role played by the pre-engineered dislocations in controlling the brittle to ductile transition at room temperature. These pre-inserted dislocations act as sources for dislocations to further multiply and move to carry the plasticity [19].

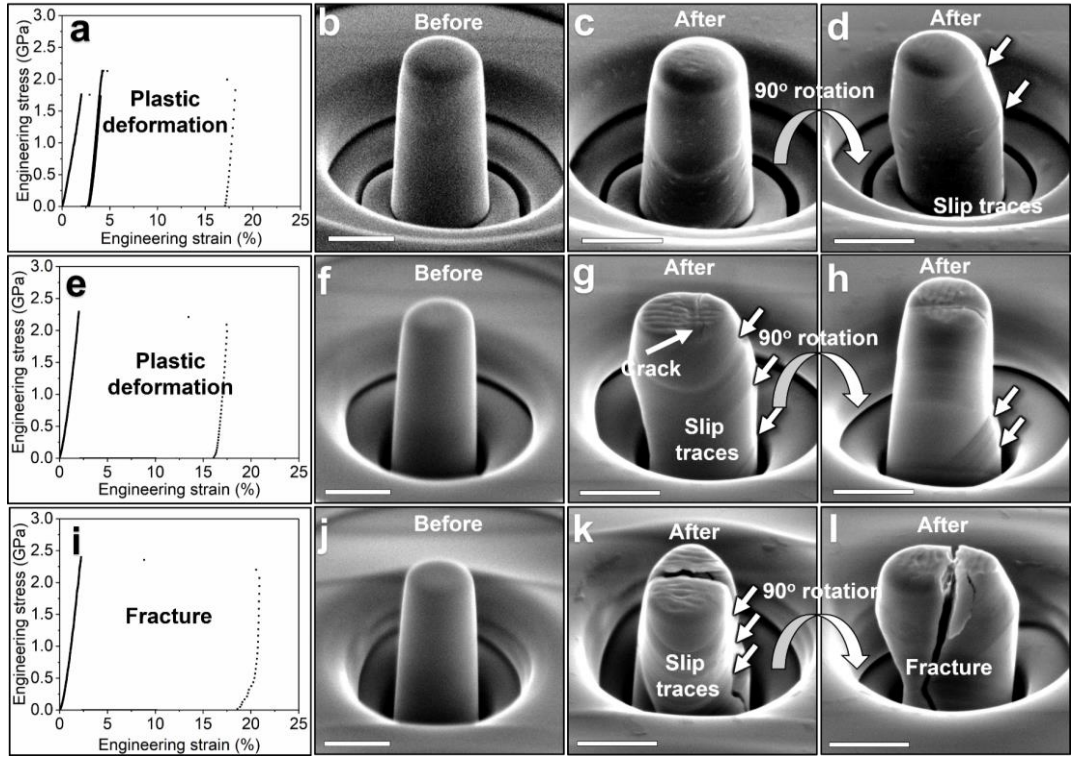

**Fig. 7.** Micropillars with unevenly distributed dislocations after 1× scratching displaying stochastic deformation behavior (either fracture or plastic deformations). (a, e, i) Representative engineering stress-strain curves. (b, f, j) SEM images of the corresponding micropillars before compression. (c-d, g-h, k-l) Surface morphologies of the compressed micropillars observed from different angles. The scale bar is 1 μm for all SEM images.

To further investigate the deformation capabilities, compression tests were performed on the 20× and 50× pillars until the samples fractured. As depicted in **Fig. 8**, both pillars achieved ~22% fracture strain, with the 50× pillar exhibiting more severe fracture damage. The 20× pillar sheared along the {110} planes, while the 50× pillar not only sheared along the {110} planes but also fractured vertically, resulting in several fragments. These observations suggest that an excessively high dislocation density may not be favorable for large plastic strain, because the strong dislocation interaction (as in the case of MgO [8]) and the pile-up of dislocations [23] can cause cracks to form.

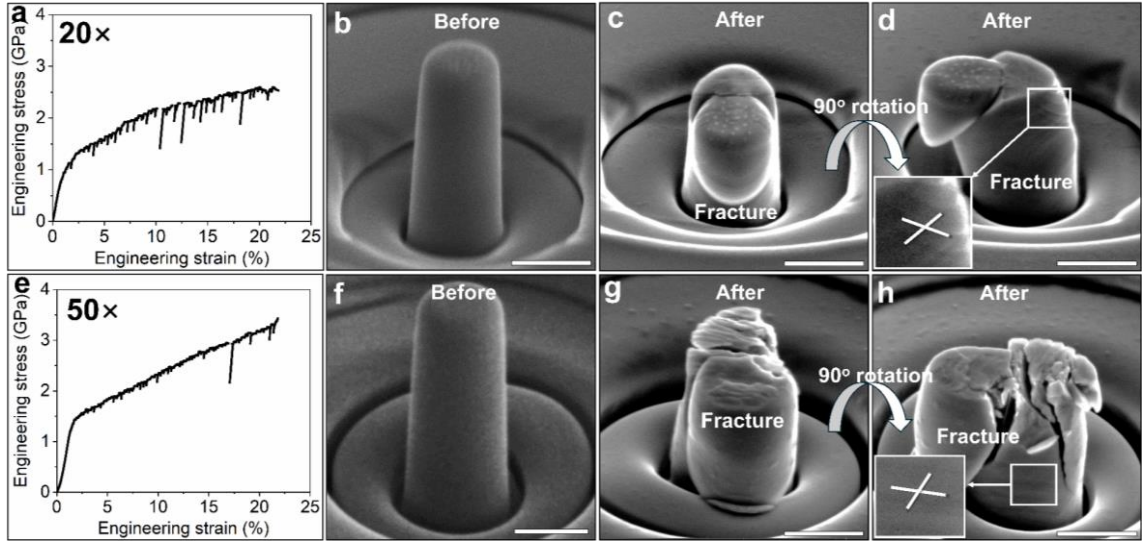

**Fig. 8.** Assessing the maximum fracture strain for micropillars with dislocation density exceeding $10^{14}$ m$^{-2}$. (a, e) Engineering stress-strain curve of the 20× and 50× micropillars both showing ~22% fracture strain. (b, f) SEM image of the two micropillars before compression. (c-d; g-h) SEM images revealing the fracture morphologies of the micropillars. The scale bar is 1 μm for all SEM images.

As an overview, we plot in **Fig. 9** the yield strength of the plastically deformed micropillars as a function of the dislocation density. For all the reference pillars and some of the 1× scratched pillars without dislocations, the samples fractured and the data is excluded in **Fig. 9**. The detailed mechanical data are summarized in **Table S2**. Note that the yield strength of the micropillars first decreased and then increased as the dislocation density increased. First of all, the drop in the yield strength in the blue background region in **Fig. 9** is attributed to the easier dislocation multiplication and motion, once the dislocation nucleation barrier is overcome [19]. Second, as the dislocation densities are over ~7.5 × $10^{13}$ m$^{-2}$, the yield strength increases (the light pink background region in **Fig. 9**). The increase of yield strength with higher dislocation densities is attributed to the dislocation forest hardening, as will be discussed in **Sec. 4** using the Taylor model.

Note that such "V-shaped" yield strength change as a function of dislocation density has been extensively discussed in metals [38] and experimentally validated in single-crystal Cu by Hildebrand [39] (although with a much lower dislocation density, ranging from $10^6$ m$^{-2}$ to $10^{10}$ m$^{-2}$). However, there exists a large gap between the experimental data and computational simulations in metals with extremely high dislocation densities beyond $10^{16}$ m$^{-2}$ [40]. In addition, because significant increase in the dislocation density in metals often leads to grain refinement and formation of (sub)grain boundaries, it can complicate the analysis of work hardening by dislocations alone. Using oxides with varying dislocation densities can be an alternative for future DDD/MD (discrete dislocation dynamics/molecular dynamics) simulations.

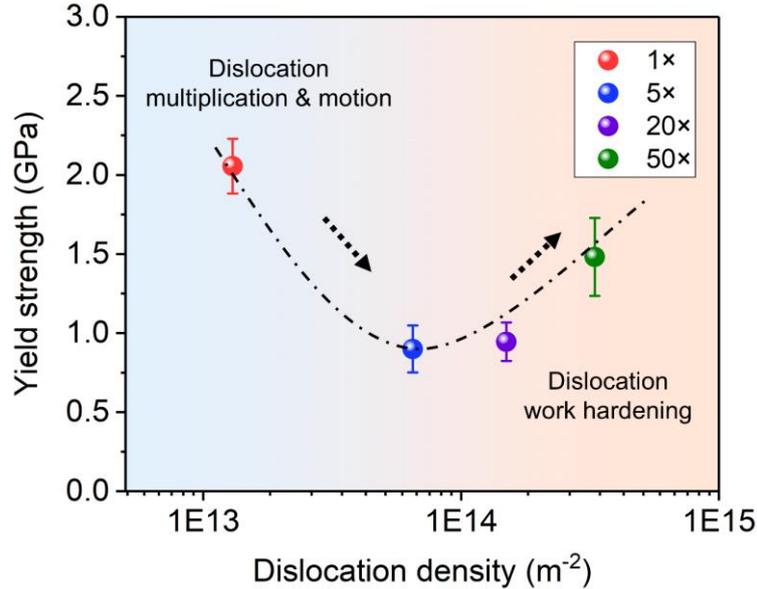

**Fig. 9** Yield strength of SrTiO$_3$ micropillars (~1 μm in diameter) as a function of the dislocation densities. The data for the reference pillars and some of the 1× scratched pillars that showed fracture behavior is excluded.

### 3.3. Dislocation structure after compression

TEM analyses were performed on two representative micropillars to examine the dislocation structure: one subjected to 5× scratching with 10% strain (**Fig. 6h**) and the other to 20× scratching with 20% strain (**Fig. 6l**). TEM lamellae were extracted from the pillar centers (see **Fig. 10**). For the 5× pillar with 10% strain, TEM images in **Figs. 10a-c** reveal 45° cross-hatched dislocations (dashed lines) with an increased density at the top of the pillar, reaching approximately $10^{14}$ m$^{-2}$. The pillar's side surface displayed a zig-zag morphology (highlighted by the white arrows in **Fig. 10a**), corresponding to the coarse slip traces after compression (**Fig. 6h**). The pillar subjected to 20× scratching with 20% strain exhibited distinct microstructural features. It has a smooth side surface (**Fig. 10d**), reflecting the finer and smoother slip bands observed in **Fig. 6l**. Two prominent high-density dislocation walls were visible, confirming the activation of slip on these planes. In the pillar's top region, dislocation walls were parallel to the sample surface (**Fig. 10e**), while in the middle region, the dislocations were more uniformly distributed (**Fig. 10f**). Viewed from the scratching direction (SD), the dislocation structures in the 20% deformed sample appeared as shorter lines and discrete dotted features, in contrast to the longer, continuous dislocations seen in the as-scratched sample (**Fig. 2**).

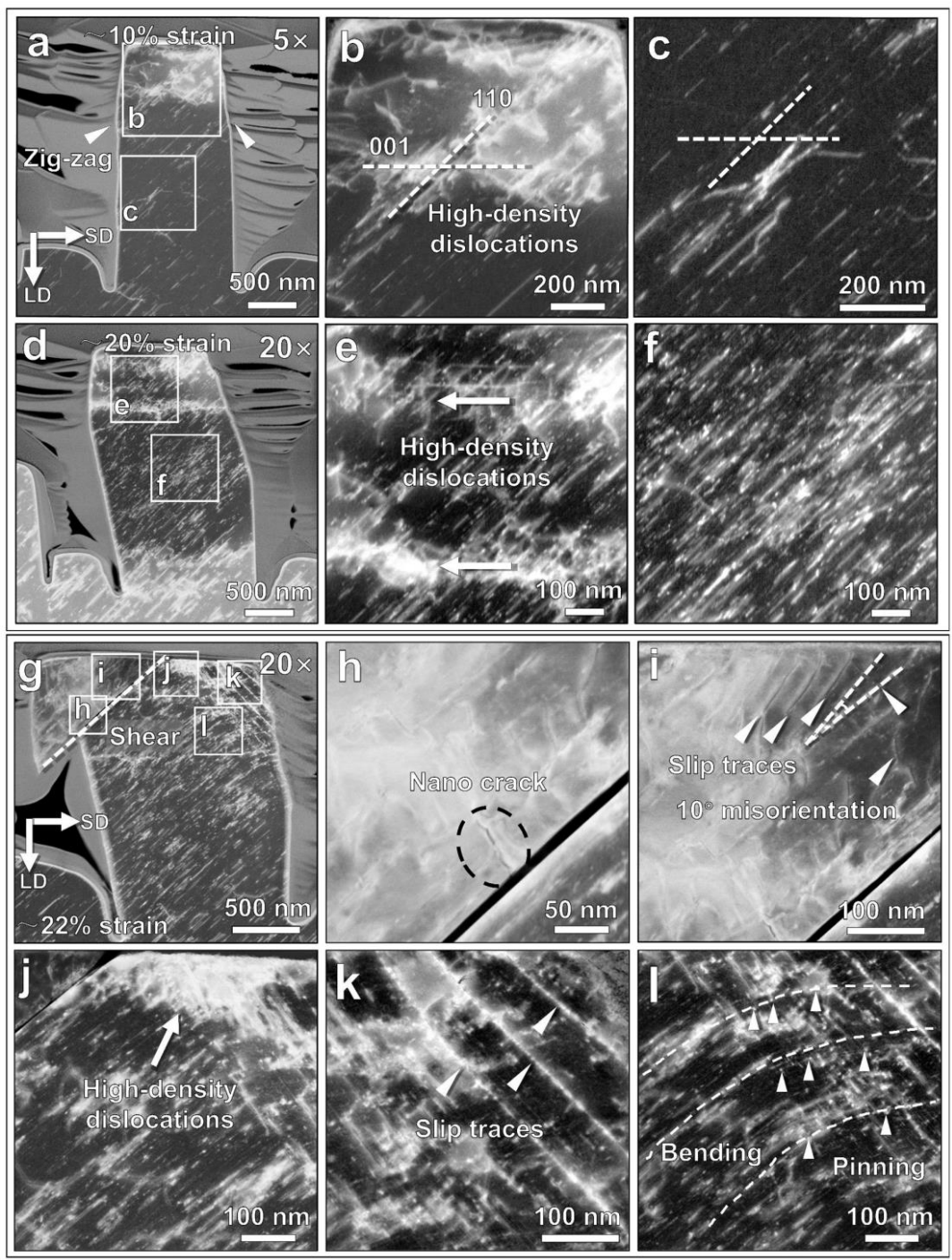

**Fig. 10** ADF-STEM micrographs of the deformed SrTiO$_3$ micropillars after compression, with TEM lamellae oriented along the plane defined by the loading direction (LD) and scratching direction (SD): (a) Dislocation structures in the 5× scratched micropillar after ~10% compression strain (refer to **Fig. 6h**). (b) Zoomed-in view of (a) showing high-density dislocations. (c) Cross-hatched dislocations were observed in the pillar's middle region. (d) Dislocation distribution in the 20× scratching micropillar after ~20% compression strain in (refer to **Fig. 6l**). (e) Magnified view of (d) highlighting two high-density dislocation walls. (f) Increased dislocation density in the pillar's middle region. ADF-STEM images of the fractured pillar (20×) after ~22% compression strain (refer to **Fig. 8d**). (g) Cross-section of the fractured micropillar. (h) Enlarged view showing a secondary

nano-crack near the fracture interface. (i) Slip traces at the top of the micropillar, with a ~10° misorientation relative to the fracture interface. (j) High-density dislocations in the pillar's top region. (k) Slip traces observed along the (011) plane. (l) Bent dislocation walls caused by the pinning effects.

To further clarify the fracture mechanism in micropillars with high dislocation densities (~$10^{14}$ m$^{-2}$), TEM analysis on the 20× pillar with 22% strain (deformed in **Figs. 8a-d**) was performed. Along the scratching direction **(Figs. 10g-l)**, the micropillar predominantly fractured on the (0-11) plane (dashed line in **Fig. 10g**). A secondary nano-crack ~80 nm in length was observed in the <110> direction (highlighted in **Fig. 10h**). At the top of the micropillar, nano-sized slip traces were observed, misoriented by ~10° misorientation relative to the primary crack. The dislocation density in this region was estimated to be greater than ~$10^{15}$ m$^{-2}$. Due to the strong interactions between dislocations, some of the dislocation walls (see **Fig. 10l**) were bent, exhibiting local pinning/locking with intersecting dislocations. These observations suggest significant dislocation forest hardening, which was evidenced by the stress-strain curves in **Fig. 6**. The interactions among dislocations contribute to pile up, which likely plays a major role in the micropillar's fracture. This TEM analysis suggests that excessively high dislocation densities can indeed impair plasticity during pillar compression, consistent with the results shown in **Fig. 8**.

## 4. Discussions

### 4.1. Size effect in micropillar compression

Unlike metals, a key distinction in the size effect of ceramics is their susceptibility to fracture with low starting dislocation densities. In dislocation-free micropillars (**Figs. 6a-d**), dislocation nucleation is the limiting process and requires much higher shear stress (e.g., for homogeneous dislocation nucleation, a shear stress of *G/2π* is required, with *G* being the shear modulus), rendering ceramic samples prone to fracture before plastic deformation can occur except for small pillars [19]. It was reported that dislocation-free SrTiO$_3$ nanopillars with diameters less than ~180 nm can undergo plastic deformation rather than immediate fracture [41]. This phenomenon, coined as size-dependent brittle-to-ductile transition, has been observed in other ceramics [19, 42] and in silicon [43].

Unlike the reported data in ceramic materials [42] focusing on pillars with very low dislocation densities, here in **Fig. 11**, we demonstrate the size effect for micropillars with dislocation density of ~1.6×$10^{14}$ m$^{-2}$ (20× scratching). For pillars with diameters of ~3 μm, the yield strength is about 0.6 GPa. The yield strength increases to around 1 GPa when the diameter is reduced to ~1 μm. An extreme case will be scaling up to bulk scale, where abundant data on single-crystal SrTiO$_3$ has been collected in literature (briefly summarized in Ref. [44], with the same loading orientation [001] as in the current work for the sake of comparison). The typically reported bulk compression tests have a yield strength of ~120 MPa [9, 44, 45]. This highlights a significant size effect, consistent with those reported in metals [46].

We also observe a transition from serrated (~1 μm pillars) to smooth (~3 μm pillars) plastic flow as reflected in the stress-strain curves in **Fig. 11**. This suggests that flow stress is influenced not only by dislocation density (as both pillars have the same dislocation density) but also by the micropillar size. Similar behavior has been reported in Cu [47], Ni [48], and LiF crystals [49], suggesting a common phenomenon across different materials. At microscopic scale, both the distribution of dislocation sources and the stress required to activate the weakest source change with the sample volume [49]. As shown in **Fig. 11**, stress drops almost diminish as the micropillar size increases to ~3 μm with the same dislocation density. This suggests a reduction in dislocation avalanche behavior due to the increased distance for dislocations to escape to the free surface and greater difficulty in dislocation emission or escape from pinned sources. Consequently, flow stress becomes smoother, and stress drops vanish [47, 50]. In the ~1 μm micropillar, minor strain bursts are observed as the applied stress is sufficient to activate internal dislocation or surface sources. When the applied load reaches the critical stress for new source multiplication, larger stress drops occur. Such clear stress drops in the smaller-sized pillars are also consistent with the hypothesis that the dislocations escaped more easily to the free surface, which may also have caused such stress drops.

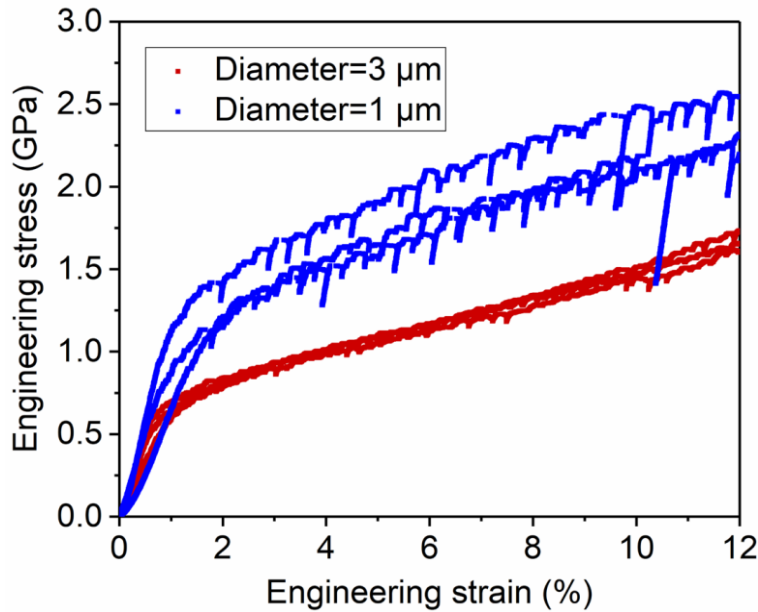

**Fig. 11.** Comparison of the engineering stress-strain curves highlighting the size effect (yield strength and stress drops) in micropillar compression with the same dislocation density (~1.3×10$^{14}$ m$^{-2}$, 20× scratching).

## 4.2. Work hardening in micropillar compression

Work hardening in ceramics has been much less discussed, primarily due to the reason that dislocations were not commonly considered relevant in most ceramics, particularly at room temperature. The current dislocation-rich samples offer the opportunity to address this point. To avoid the stochastic behavior in the 1× scratching, we focus on the three conditions with scratching passes on 5×, 20×, and 50×. To evaluate the yield strength as a function of dislocation density during micropillar compression, for simplicity, we applied the Taylor hardening relationship developed particularly for pillar compression. The Taylor hardening formula is given by [51, 52]:

$$\tau = \tau_0 + \alpha\mu b\sqrt{\rho} + \frac{k\mu b}{\bar{\lambda}} \tag{1}$$

where $\tau$ is the yield strength, $\tau_0$ the intrinsic strength of the materials (related to the lattice friction stress and long-range internal stress), $\alpha$ a constant term factor, $\mu$ the shear modulus (~107 GPa for SrTiO$_3$ at room temperature) [53], $b$ the length of the Burgers vector (~0.55 nm for SrTiO$_3$), $\rho$ the dislocation density, $k$ is a constant, and $\bar{\lambda}$ is the average source length. Here, we focus on the net increase in the yield strength caused by the increase in dislocation density, namely:

$$\Delta\tau = \tau - \tau_0 - \frac{k\mu b}{\bar{\lambda}} = \alpha\mu b\sqrt{\rho} \tag{2}$$

As illustrated in **Fig. 12**, the Taylor model can capture the net increase in the yield strength (with fitted value $\alpha = 1.1$, black dotted line) for the case of 5×, 20×, and 50× (corresponding to dislocation density of ~7.5×10$^{13}$ m$^{-2}$, ~1.6×10$^{14}$ m$^{-2}$, and ~3.3×10$^{14}$ m$^{-2}$). The individual values of the net increase in the yield strengths are listed in **Table S3**. The fitted value of $\alpha = 1.1$ in this study implies the yield strengths of the micropillars are proportionally to the square of the dislocation density.

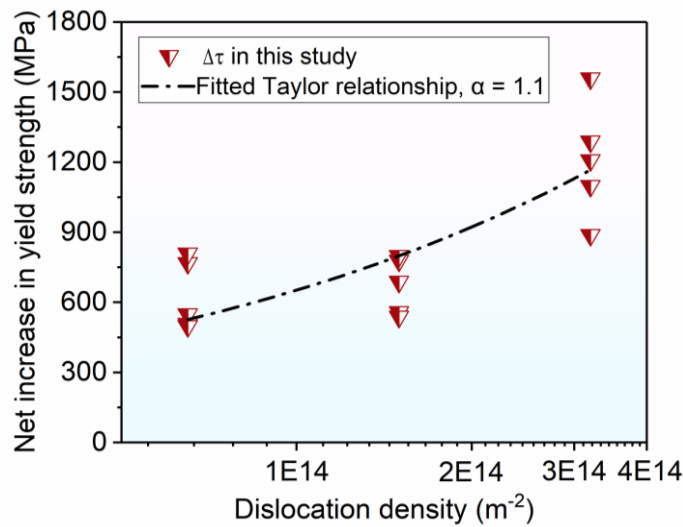

**Fig. 12**. Estimated net increase in yield strength as a function of the dislocation density, and fitted Taylor hardening relationship (black dotted line) indicates the Taylor term value α=1.1.

## 5. Conclusion

Using room-temperature cyclic scratching on single-crystal SrTiO$_3$, we generated plastic zones in the range of 100 μm in widths and depths, with dislocation densities ranging from ~$10^{10}$ m$^{-2}$ to ~$10^{14}$ m$^{-2}$. This approach facilitates the investigation of dislocation-dependent strength of oxides. We find that room-temperature brittle to ductile transition in micropillar compression of SrTiO$_3$ is mitigated by the pre-engineered dislocations, which circumvent dislocation nucleation and promote plasticity. In the plastic deformation regime, the yield strength decreases first and then increases, analogous to metals. This corresponds to the dominating dislocation activities from multiplication to impeded mobility caused by work hardening, also reflected in the size effect in SrTiO$_3$ pillar compression. Importantly, high dislocation densities (~$10^{14}$ m$^{-2}$ and beyond) can be counter-effective for dislocation plasticity, as fracture of the samples can be induced by stronger dislocation interactions and pile-up. These findings highlight the potential of dislocation engineering as a useful approach for understanding the mechanical properties of ceramics.

**Acknowledgement:** W. Lu acknowledges the support by Shenzhen Science and Technology Program (grant no. JCYJ20230807093416034), the Open Fund of the Microscopy Science and Technology-Songshan Lake Science City (grant no. 202401204), National Natural Science Foundation of China (grant no. 52371110) and Guangdong Basic and Applied Basic Research Foundation (grant no. 2023A1515011510). X. Fang thanks the support by the European Union (ERC Starting Grant, Project MECERDIS, grant no. 101076167). Views and opinions expressed are however those of the authors only and do not necessarily reflect those of the European Union or the European Research Council. Neither the European Union nor the granting authority can be held responsible for them. The authors acknowledge the use of the facilities at the Southern University of Science and Technology Core Research Facility.

**Author contributions:** W. Lu and X. Fang conceived the idea, designed the experiments, and supervised the project. J. Zhang performed the experimental tests and collected the data. J. Zhang wrote the first draft. All authors discussed, interpreted the data, and revised the manuscript.

**Conflict of Interest:** Authors declare that they have no competing interests.

**Data accessibility:** The data for this study is available in the main text or the supplementary materials. The raw data can be provided by the corresponding authors upon reasonable request.

**References**

[1]  M.A. Keane, Ceramics for catalysis, J. Mater. Sci. 38 (2003) 4661-4675.

[2]  R.C. Buchanan, Ceramic materials for electronics (3rd edition), CRC Press, 2018.

[3]  H. Palneedi, M. Peddigari, G.T. Hwang, D.Y. Jeong, J. Ryu, High-performance dielectric ceramic films for energy storage capacitors: progress and outlook, Adv. Funct. Mater. 28 (2018) 1803665.

[4]  J. Zhang, G. Liu, W. Cui, Y. Ge, S. Du, Y. Gao, Y. Zhang, F. Li, Z. Chen, S. Du, K. Chen, Plastic deformation in silicon nitride ceramics via bond switching at coherent interfaces, Science 378 (2022) 371–376.

[5]  L.R. Dong, J. Zhang, Y.Z. Li, Y.X. Gao, M. Wang, M.X. Huang, J.S. Wang, K.X. Chen, Borrowed dislocations for ductility in ceramics, Science 385 (2024) 422–427.

[6]  I.W. Chen, Implications of transformation plasticity in $ZrO_2$-containing ceramics: II, elastic-plastic indentation, J. Am. Ceram. Soc. 69 (1986) 189–194.

[7]  J.J. Gilman, W.G. Johnston, Dislocations in lithium fluoride crystals, Solid state phys. 13 (1962) 147–222.

[8]  A.S. Argon, E. Orowan, Plastic deformation in MgO single crystals, Philos. Mag. 9 (1964) 1003–1021.

[9]  D. Brunner, S. Taeri-Baghbadrani, W. Sigle, M. Rühle, Surprising results of a study on the plasticity in strontium titanate, J. Am. Ceram. Soc. 84 (2001) 1161–1163.

[10] A.F. Mark, M. Castillo-Rodriguez, W. Sigle, Unexpected plasticity of potassium niobate during compression between room temperature and 900 °C, J. Eur. Ceram. Soc. 36 (2016) 2781–2793.

[11] X. Fang, J. Zhang, A. Frisch, O. Preuß, C. Okafor, M. Setvin, W. Lu, Room-temperature bulk plasticity in $KTaO_3$ and tunable dislocation densities, J. Am. Ceram. Soc.107 (2024) 7054–7061.

[12] X. Fang, Mechanical tailoring of dislocations in ceramics at room temperature: A perspective, J. Am. Ceram. Soc. 107.3 (2024) 1425–1447.

[13] C. Shen, J. Li, T. Niu, J. Cho, Z. Shang, Y. Zhang, A. Shang, B. Yang, K. Xu, R.E. García, H. Wang, X. Zhang, Achieving room temperature plasticity in brittle ceramics through elevated temperature preloading, Sci. Adv. 10 (2024) eadj4079.

[14] K.H. Yang, N.J. Ho, H.Y. Lu, Plastic deformation of <001> single-crystal $SrTiO_3$ by compression at room temperature, J. Am. Ceram. Soc. 94 (2011) 3104–3111.

[15] J. Nishigaki, K. Kuroda, H. Saka, Electron microscopy of dislocation structures in $SrTiO_3$ deformed at high temperatures, Phys. Status Solidi (a) 128 (1991) 319–336.

[16] L. Jin, X. Guo, C.L. Jia, TEM study of <110>-type 35.26° dislocations specially induced by polishing of $SrTiO_3$ single crystals, Ultramicroscopy 134 (2013) 77–85.

[17] X. Fang, H. Bishara, K. Ding, H. Tsybenko, L. Porz, M. Höfling, E. Bruder, Y. Li, G. Dehm, K. Durst, Nanoindentation pop-in in oxides at room temperature: Dislocation activation or crack formation?, J. Am. Ceram. Soc. 104 (2021) 4728–4741.

[18] C. Okafor, K. Ding, X. Zhou, K. Durst, J. Rödel, X. Fang, Mechanical tailoring of dislocation densities in SrTiO$_3$ at room temperature, J. Am. Ceram. Soc. 105 (2022) 2399–2402.

[19] X. Fang, W. Lu, J. Zhang, C. Minnert, J. Hou, S. Bruns, U. Kunz, A. Nakamura, K. Durst, J. Rödel, Harvesting room-temperature plasticity in ceramics by mechanically seeded dislocations, Mater. Today. 82 (2025) 81–91.

[20] T. Scholz, K.K. McLaughlin, F. Giuliani, W.J. Clegg, F.J. Espinoza-Beltrán, M. V. Swain, G.A. Schneider, Nanoindentation initiated dislocations in barium titanate (BaTiO$_3$), Appl. Phys. Lett. 91 (2007) 062903.

[21] Y.J. Kim, I.C. Choi, J.A. Lee, M.Y. Seok, J.I. Jang, Strain-dependent transition of time-dependent deformation mechanism in single-crystal ZnO evaluated by spherical nanoindentation, Philos. Mag. 95 (2015) 1896–1906.

[22] A. Montagne, V. Audurier, C. Tromas, Influence of pre-existing dislocations on the pop-in phenomenon during nanoindentation in MgO, Acta Mater. 61 (2013) 4778–4786.

[23] X. Fang, K. Ding, C. Minnert, A. Nakamura, K. Durst, Dislocation-based crack initiation and propagation in single-crystal SrTiO$_3$, J. Mater. Sci. 56 (2021) 5479–5492.

[24] F. Javaid, A. Stukowski, K. Durst, 3D Dislocation structure evolution in strontium titanate: Spherical indentation experiments and MD simulations, J. Am. Ceram. Soc. 100 (2017) 1134–1145.

[25] O. Preuß, E. Bruder, W. Lu, F. Zhuo, C. Minnert, J. Zhang, J. Rödel, X. Fang, Dislocation toughening in single-crystal KNbO$_3$, J. Am. Ceram. Soc. 106 (2023) 4371–4381.

[26] O. Preuß, E. Bruder, J. Zhang, W. Lu, J. Rödel, X. Fang, Damage-tolerant oxides by imprint of an ultra-high dislocation density, J. Eur. Ceram. Soc. 45 (2025) 116969.

[27] W.D. Nix, H. Gao, Indentation size effects in crystalline materials: A law for strain gradient plasticity, J. Mech. Phys. Solids. 46 (1998) 411–425.

[28] X. Fang, O. Preuß, P. Breckner, J. Zhang, W. Lu, Engineering dislocation-rich plastic zones in ceramics via room-temperature scratching, J. Am. Ceram. Soc. 106 (2023) 4540–4545.

[29] Y. Meng, X. Ju, X. Yang, The measurement of the dislocation density using TEM, Mater. Charact. 175 (2021) 111065.

[30] H. Mughrabi, Dislocation wall and cell structures and long-range internal stresses in deformed metal crystals, Acta Metall. 31 (1983) 1367–1379.

[31] L. Porz, A.J. Klomp, X. Fang, N. Li, C. Yildirim, C. Detlefs, E. Bruder, M. Höfling, W. Rheinheimer, E.A. Patterson, P. Gao, K. Durst, A. Nakamura, K. Albe, H. Simons, J. Rödel, L. Porz, Dislocation-toughened ceramics, Mater. Horiz. 8 (2021) 1528–1537.

[32] P. Hirsch, D. Cockayne, J. Spence, M. Whelan, 50 Years of TEM of dislocations: Past, present and future, Philos. Mag. 86 (2006) 4519–4528.

[33] D.J. Bacon, D. Hull, Introduction to dislocations, Butterworth-Heinemann, (2011) pp. 238-243.

[34] C. Greiner, Z. Liu, L. Strassberger, P. Gumbsch, Sequence of Stages in the Microstructure Evolution in Copper under Mild Reciprocating Tribological Loading, ACS Appl. Mater. Interfaces 8 (2016) 15809–15819.

[35] R. Niu, X. An, L. Li, Z. Zhang, Y.W. Mai, X. Liao, Mechanical properties and deformation behaviours of submicron-sized Cu–Al single crystals, Acta Mater. 223 (2022) 117460.

[36] S.H. Li, Y. Zhao, J. Radhakrishnan, U. Ramamurty, A micropillar compression investigation into the plastic flow properties of additively manufactured alloys, Acta Mater. 240 (2022) 118290.

[37] T. Hu, L. Jiang, H. Yang, K. Ma, T.D. Topping, J. Yee, M. Li, A.K. Mukherjee, J.M. Schoenung, E.J. Lavernia, Stabilized plasticity in ultrahigh strength, submicron Al crystals, Acta Mater. 94 (2015) 46–58.

[38] L. Johnsonts, M.F. Ashbyt, The stress at which dislocations multiply in well-annealed metal crystals, Acta Metall. 16.2 (1968) 219–225.

[39] H. Hildebrand, The effect of the initial dislocation density on dislocation multiplication and work-hardening characteristics of copper single crystals, Phys. Status Solidi (a) 12 (1972) 239–249.

[40] H. Fan, Q. Wang, J.A. El-Awady, D. Raabe, M. Zaiser, Strain rate dependency of dislocation plasticity, Nat. Commun. 12 (2021) 1845.

[41] Y. Liu, X. Cui, R. Niu, S. Zhang, X. Liao, S.D. Moss, P. Finkel, M. Garbrecht, S.P. Ringer, J.M. Cairney, Giant room temperature compression and bending in ferroelectric oxide pillars, Nat. Commun. 13 (2022) 335.

[42] S. Korte-Kerzel, Microcompression of brittle and anisotropic crystals: Recent advances and current challenges in studying plasticity in hard materials, MRS Commun. 7 (2017) 109–120.

[43] F. Östlund, K. Rzepiejewska-Malyska, K. Leifer, L.M. Hale, Y. Tang, R. Ballarini, W.W. Gerberich, J. Michler, Brittle-to-ductile transition in uniaxial compression of silicon pillars at room temperature, Adv. Funct. Mater. 19 (2009) 2439–2444.

[44] S. Stich, K. Ding, Q.K. Muhammad, L. Porz, C. Minnert, W. Rheinheimer, K. Durst, J. Rödel, T. Frömling, X. Fang, Room-temperature dislocation plasticity in $SrTiO_3$ tuned by defect chemistry, J. Am. Ceram. Soc. 105 (2022) 1318–1329.

[45] D. Brunner, Low-temperature plasticity and flow-stress behaviour of strontium titanate single crystals, Acta Mater. 54 (2006) 4999–5011.

[46] M.D. Uchic, D.M. Dimiduk, J.N. Florando, W.D. Nix, Sample dimensions influence strength and crystal plasticity, Science 305.5686 (2004) 986-989.

[47] J.Y. Zhang, X. Liang, P. Zhang, K. Wu, G. Liu, J. Sun, Emergence of external size effects in the bulk-scale polycrystal to small-scale single-crystal transition: A maximum in the strength and strain-rate sensitivity of multicrystalline Cu micropillars, Acta Mater. 66 (2014) 302–316.

[48] C.P. Frick, B.G. Clark, S. Orso, A.S. Schneider, E. Arzt, Size effect on strength and strain hardening of small-scale [111] nickel compression pillars, Mater. Sci. Eng. A 489 (2008) 319–329.

[49] E.M. Nadgorny, D.M. Dimiduk, M.D. Uchic, Size effects in LiF micron-scale single crystals of low dislocation density, J. Mater. Res. 23.11 (2008) 2829–2835.

[50] M. Zaiser, P. Moretti, Fluctuation phenomena in crystal plasticity - A continuum model, J. Stat. Mech. Theory Exp. 2005.08 (2005) 79–97.

[51] Y. Cui, Z. Liu, Z. Zhuang, Quantitative investigations on dislocation based discrete-continuous model of crystal plasticity at submicron scale, Int. J. Plast. 69 (2015) 54–72.

[52] T.A. Parthasarathy, S.I. Rao, D.M. Dimiduk, M.D. Uchic, D.R. Trinkle, Contribution to size effect of yield strength from the stochastics of dislocation source lengths in finite samples, Scr. Mater. 56 (2007) 313–316.

[53] F. Javaid, K.E. Johanns, E.A. Patterson, K. Durst, Temperature dependence of indentation size effect, dislocation pile-ups, and lattice friction in (001) strontium titanate, J. Am. Ceram. Soc. 101 (2018) 356–364.